\documentclass[bibyear]{aa}

\usepackage{graphicx}
\usepackage{amsmath}
\usepackage{epsfig}
\usepackage{txfonts}
\usepackage{url}

\def\teff{$T_{\rm eff}$}
\def\teffp{$T_{\rm eff}^{\rm pole}$}
\def\logg{$\log g$}
\def\loggp{$\log g^{\rm pole}$}
\def\kms{km\,s$^{-1}$}

\newcommand\he[2]{He\,\textsc{#1}\,$\lambda$\,{#2}}
\newcommand\si[2]{Si\,\textsc{#1}\,$\lambda$\,{#2}}

\begin{document}

   \title{Spectral modelling of the Alpha Virginis (\textit {Spica}) binary system}

   \author{M. Palate\inst{1}
          \and
          G. Koenigsberger\inst{2}
          \and
          G. Rauw\inst{1}
          \and
          D. Harrington\inst{3}
          \and
          E. Moreno\inst{4}
          }

   \institute{Institut d'Astrophysique et de G\'eophysique, 
   Universit\'e de Li\`ege, B\^at. B5c, All\'ee du 6 Ao\^ut 17, 4000 Li\`ege, Belgium\\
             \email{palate@astro.ulg.ac.be}
             \and
             Instituto de Ciencias F\'{\i}sicas, Universidad Nacional Aut\'onoma de M\'exico, 
             Cuernavaca, Morelos 62210, M\'exico\\
						 Currently at Instituto de Astronom\'{\i}a, UNAM, Apdo. Postal 70-264, M\'exico D.F.\\
							\email{gloria@astro.unam.mx}
             \and
						 Institute for Astronomy, University of Hawaii, 2680 Woodlawn Drive, 
             Honolulu, HI, 96822, USA\\
                                                        \email{dmh@ifa.hawaii.edu}
             \and
             Instituto de Astronom\'{\i}a, Universidad Nacional Aut\'onoma de M\'exico, 
             M\'exico, D.F., M\'exico\\
                                                        \email{edmundo@astro.unam.mx}
             }

   \date{Received 17 May 2013 / Accepted 20 June 2013}

	\abstract
{The technique of matching synthetic spectra computed with theoretical stellar atmosphere models to the observations is widely used in deriving fundamental parameters of massive stars. When applied to binaries, however, these models generally neglect the interaction effects present in these systems.}
{The aim of this paper is to explore the uncertainties in binary stellar parameters that are derived from single-star models.}
{Synthetic spectra that include the tidal perturbations and irradiation effects are computed for the binary system $\alpha$ Virginis (\textit {Spica}) using our recently-developed CoMBiSpeC model. The synthetic spectra are compared to $S/N\sim$2000 observations and optimum values of \teff\ and \logg\ are derived.}
{The binary interactions have only a small effect on the strength of the photospheric absorption lines in \textit {Spica} ($<2$\% for the primary and $<4$\% for the secondary). These differences are comparable to the uncertainties inherent to the process of matching synthetic spectra to the observations and thus the derived values of \teff\ and \logg\ are unaffected by the binary perturbations. On the other hand, the interactions do produce significant phase-dependent line profile variations in the primary star, leading to  systematic distortions in the shape of its radial velocity curve. Migrating sub-features (``bumps'') are predicted by our model to be present in the same photospheric lines as observed, and their appearance does not require any a priori assumptions regarding non-radial pulsation modes.  Matching the strength  of lines in which the most prominent ``bumps'' occur requires synthetic spectra computed with larger ``microturbulence'' than that required by other lines.}
{}


   \maketitle

\section{Introduction}


Massive stars play a key role in the evolution of galaxies through their high luminosity, powerful winds, heavy element enrichment of the ISM, and the explosions when they end their lives as supernovae. The importance of determining their fundamental parameters cannot be overstated, and considerable effort has been invested in obtaining high quality observations and applying theoretical stellar structure models to establish these parameters (see for example, Evans et al. \cite{Evans}; Martins \cite{Martins}; Massey et al. \cite{Massey09, Massey12}). Although significant progress has been made, there are still some important gaps in our understanding of the physical processes that govern the structure and evolution of massive stars. One of these gaps involves the effects caused by a binary companion on the emergent spectrum, from which the fundamental parameters are generally derived. 

Binary systems provide the only direct means of determining the  masses of stars and thus they are used to test the models of stellar structure which are then applied to single stars. The usual method for obtaining the binary star parameters is through the use of the radial velocity (RV) curves  and  model atmosphere codes. This yields a lower limit to the masses of the two stars, $m~\sin^3i$, the stellar radii $R$, effective temperature \teff, surface gravitational acceleration \logg, and metallicity (see, for example, Massey et al. \cite{Massey12}).

The majority of stellar atmosphere  models  have been developed for single stars. Thus, one of the important questions that arises  concerns the extent to which they can be used to properly model binary stars since the latter are subject to a variety of interaction effects. For example, gravity darkening and irradiation of the hemisphere facing the companion are expected to lead to different \teff\ values over the surface, with analogous differences in \logg\ values due to the tidal distortion. In addition, non-synchronous rotation and orbital eccentricity induce oscillations on the stellar surface  that may lead to photospheric absorption line profiles that are significantly different from those that are predicted by single-star atmospheric models (Vogt \& Penrod \cite{Vogt}).  It is possible that the temporal variability of these line profiles can have a significant impact on the determination of the fundamental parameters through the distortion of the radial velocity curves (Koenigsberger et al. \cite{Koenigsberger}). 

We have developed a method for computing the spectra of binary systems which takes into account interaction effects. In a first stage, the code was designed for circular massive binary systems (Palate \& Rauw \cite{Palate}) in which the  distorted shape of the stellar surfaces due to the gravitational interaction is calculated and then  the emergent spectra at different orbital phases are computed. In a second paper the method was extended to include eccentric binaries and to incorporate radiation pressure effects (Palate et al. \cite{PRKM}). This new version of the CoMBiSpeC (Code of Massive Binary Spectral Computation) now allows spectral computation of massive binary systems in general. In this paper we use this code to address the question of the extent to which the binary interactions may affect the outcome of the standard procedure of matching stellar atmosphere models to observations in order to derive values of \teff\ and \logg. This exploratory investigation will focus on the nearby and well-studied binary system $\alpha$ Virginis (= {\it Spica} = HD\,116658). 

In this paper we use local line-profiles obtained from a full radiative transfer computation using the standard model atmosphere code TLUSTY. This enables us to analyse the effects of the binary interaction on the determination of \teff\ and \logg, in addition to the
line-profile variability. In Section 2 we provide a brief review of {\it Spica}'s properties; in Section 3 we describe the procedure that was followed to produce the synthetic spectra, in Section 4 we present the results, and in Section 5 the conclusion.

\section{The \textit {Spica} binary system}

{\it Spica} is a double-lined, short-period ($\sim$4 days) spectroscopic binary in an eccentric orbit. The primary component is classified as B1.5 IV-V and based on early observations that disclosed a 4.17 hr spectroscopic and photometric period, it was believed to be a $\beta$ Cephei-type star (Shobbrook et al. \cite{Shobbrook1,Shobbrook2}). However, the photometric variations seem to have vanished in 1970-1971 (Smith 1985a). On the other hand, the spectroscopic line-profile variations have persisted (Smak \cite{Smak}; Smith \cite{Smitha}, \cite{Smithb}; Riddle \cite{Riddle}; Harrington \cite{Harrington}). They are commonly described in terms of travelling ``bumps'' that migrate from the blue to the red wing of the weak photospheric absorption lines, but also include variations in the slope of the line wings.  A periodicity of 6.5 and 3.2 hrs has been associated with these variations (Smith \cite{Smitha}).

{\it Spica} was observed interferometrically by Herbison-Evans et al. (\cite{Herbison}), allowing a direct determination of orbital separation, stellar radius of the primary ($R_1=8.1\pm0.5 R_{\sun}$) and orbital inclination ($i=65.9\degr\pm1.8\degr$), which when combined with radial velocity (RV) curves yield the masses, $M_1=10.9\pm0.9 M_{\sun}$ and $M_2=6.8\pm0.8 M_{\sun}$.  More recently, it was observed with the CHARA and SUSI interferometric arrays by Aufdenberg (\cite{Aufdenberg}) and collaborators. The analysis followed that of Herbison-Evans et al. (\cite{Herbison}) except that the stellar disks were no longer assumed to be of uniform brightness but were assumed to be rotationally  and tidally distorted.  The results of this analysis were kindly provided to us by J. Aufdenberg (2008, private communication) and consist of slightly different values for the stellar and orbital parameters. They are listed in Table~\ref{spicaprop} together with those of Herbison-Evans et al. (\cite{Herbison}).  

\begin{table}[!h,!t,!b]
	\begin{center}
	\caption{\textit {Spica} Parameters \label{spicaprop}}
		\begin{tabular}{lcc}
		\hline
		\hline
		{\bf Parameter}       & {\bf H-E}$^{(a)}$   &{\bf Aufdenberg et al.}      \\
		\hline
		\hline
		Spectrum                &B1.5 IV-V + B3V         & B0.5 III-IV + B2.5-B3 V \\
		$m_1$ ($M_{\sun}$)       & 10.9$\pm$0.9          & 10.25$\pm$0.68    \\
		$m_2$ ($M_{\sun}$)       & 6.8$\pm$0.7           & 6.97$\pm$0.46      \\
		$R_1$ ($R_{\sun}$)       & 8.1$\pm$0.5           & 7.40$\pm$0.57  \\        
		$R_2$ ($R_{\sun}$)       &  ---                  & 3.64$\pm$0.28  \\
		P (day)                & 4.014597              & 4.0145898        \\
		T$_0$ (JD)              & 2440678.09            & 2440678.09       \\
		$e$                       & 0.146                 & 0.067$\pm$0.014$^{(b)}$  \\
		$i$ ($\degr$)               & 66$\pm$2              & 54$\pm$6        \\         
		$\omega$ ($\degr$) at T$_0$  & 138 $\pm15$           & 140 $\pm$10      \\
		Apsidal Period (yrs)    & 124$\pm$11            & 130$\pm$8        \\
		${\mathrm v}_1 \sin i$ (\kms) & 161$\pm$2$^{(c)}$     & 161$\pm$2$^{(c)}$ \\
		${\mathrm v}_{rot1}$ (\kms) & 176$\pm$5             & 199$\pm$5        \\         
		${\mathrm v}_2 \sin i$ (\kms)& 70$\pm$5$^{(c)}$      & 70$\pm$5$^{(c)}$ \\
		${\mathrm v}_{rot2}$ (\kms)& 77$\pm$6              & 87$\pm$6        \\
		$\beta_0(m_1)^{(d)}$     & 1.3                   & 1.88$\pm$0.19    \\        
		$\beta_0(m_2)$           &  ---                  & 1.67$\pm$0.5     \\           
		\hline
		\hline
		\end{tabular}
	\end{center}
	\tablefoot{$^\text{(a)}$ Herbison-Evans et al. (\cite{Herbison}). $^\text{(b)}$ From Riddle (\cite{Riddle}). $^\text{(c)}$ From Smith (1985a). $^\text{(d)}$ The $\beta_0$ parameter is the ratio of the rotation and orbital angular velocities at periastron and can be expressed as: $\beta_0$ = 0.02 $\frac{P~v_{rot}}{R} \frac{(1-e)^{3/2}}{(1+e)^{1/2}}$, where $v_{rot}$ is the rigid body rotation velocity (in \kms), $R$ is the equilibrium radius (in $R_{\sun}$), and $e$ is the eccentricity.}
\end{table}

The stellar rotation velocities, $v_{rot}$, were derived by Smith (\cite{Smitha}), who modelled the line-profile variability under the assumption that it could be described in terms of high-order non-radial pulsation modes. To derive the order of the modes, Smith (1985a) used a trial and error profile-fitting method in which he included the presence of travelling ``bumps'' and tested combinations of rotational, macroturbulent and pulsational velocities and periods. The rotation velocities he derived are ${\mathrm v}_1 \sin i=161\pm2$ \kms and ${\mathrm v}_2\sin i =70\pm 5$ \kms.

Harrington et al. (\cite{Harrington}) adopted the Aufdenberg et al. (2008) stellar and orbital parameters and Smith's (\cite{Smitha}) values for $v_{rot}$, and performed an {\it ab initio} calculation of the line profiles at several orbital phases in order to study the  variability that is caused by the response of the star to  tidal perturbations. The calculation involves the solution of the equations of motion of the surface elements in the presence of gravitational, Coriolis, centrifugal, viscous and gas pressure forces, and the projection of the resulting velocity field along the line of sight to the observer in order to compute photospheric absorption lines in the observer's reference frame. Harrington et al. (\cite{Harrington}) were able to reproduce the general trends in the line-profile variability and, in particular, the observed relative strength and number of ``blue'' to ``red'' migrating ``bumps''. These ``bumps'' were found to arise in what Harrington et al. describe as ``tidal flows'', a concept that differs from the non-radial pulsation representation in that the travelling waves on the stellar surface are a consequence entirely of the response of this surface to the forcing and restoring agents, the interior structure of the star playing no role. The line profile calculation performed by Harrington et al. (\cite{Harrington}) was done for an arbitrary absorption line assuming  that the local line profile at each location on the stellar surface has a Gaussian shape. Thus, their study was limited to the analysis of line-profile variability alone, and no comparison of the effects on different lines (particularly those used for temperature and gravity diagnostics) was possible.  

\section{Method of analysis}

Synthetic spectra were produced for a binary system under the assumptions of a) no interaction effects, and b) tidal and irradiation effects. These spectra will henceforth be alluded to as ``unperturbed'' and ``perturbed'', respectively. The detailed procedure for generating the synthetic spectra is described below.

Rotating and binary stars are deformed from spherical symmetry and thus the gravity darkening effects\footnote{As in Palate et al. (\cite{PRKM}) we used a gravity darkening parameter equal to $0.25$.} lead to a non-uniform value of effective temperature and gravity over the stellar surface. When synthetic spectra are produced from model atmosphere grids, they are characterized by the values of effective temperature and surface gravity at the pole. Thus, we refer to our models with the values of \teffp\ and \loggp.

The synthetic spectra were compared with observational data that were obtained on 2008 March 15--28 at the Canada France Hawaii (CFHT) 3.6m telescope with the ESPaDOns spectropolarimeter, which are thoroughly discussed in Harrington et al. (\cite{Harrington}). They consist of high resolution spectra ($\text{R} = 68000$) in the wavelength range $\lambda\lambda$\,3700-9200 \AA\ with typical $S/N\sim2000$.

The synthetic spectra were produced as follows:

 {\it Step 1}: The stellar surface deformation and velocity field of both stars in the binary system are computed using the TIDES code.   This requires prior knowledge of the stellar masses, radii and rotation velocity, as well as the full set of orbital parameters.
The masses and radii are those given by Aufdenberg et al. (2008) taking into account the uncertainties quoted by these authors, the eccentricity is from Riddle (\cite{Riddle}), and the longitude of periastron $\omega_{per}=255\degr$ is from Harrington et al. (\cite{Harrington}). The output consists of displacements and velocities (both radial and azimuthal) for each surface element as a function of time.  A full description of the TIDES calculation is provided in Moreno et al. (\cite{Moreno}) and a detailed description of the calculations that were performed for \textit {Spica} can be found in Harrington et al. (\cite{Harrington}). 

{\it Step 2}: We use the grid of stellar atmosphere models computed with TLUSTY (Lanz \& Hubeny \cite{Lanz}) to produce emergent flux spectra with microturbulent speeds in the range ${\mathrm v}_{turb}=2-15$ \kms using the routine SYNSPEC49\footnote{The grid of models is available at \url{http://nova.astro.umd.edu/Tlusty2002/tlusty-frames-BS06.html} and the latest version of SYNSPEC was obtained from \url{http://nova.astro.umd.edu/Synspec49/synspec.html}}. The set of spectra for the primary star in \textit {Spica} covers \teff\,$=22000$ to 25000 K and \logg\,$=3.75$ to 4.00, and for the secondary the corresponding ranges are 18000--22000\,K and \logg\,$=4.00$ to 4.25. The metallicity of these models is Solar and the SYNSPEC49 calculation was performed using the NLTE option\footnote{It is important to note that the NLTE calculation is actually only performed on a limited number of the lines (mainly H, He and C N O in later stages). Lanz \& Hubeny (\cite{Lanz}) argue that NLTE effects in the other lines are small.}.

{\it Step 3}: CoMBiSpeC is used to compute the temperature and gravity distributions and then is used to linearly interpolate and Doppler-shift the spectra obtained from the SYNSPEC49 calculation to obtain the emergent spectrum for each surface element of the star (i.e., the ``local'' line profile) and then the  final synthetic spectrum is produced by integration over the entire stellar surface. The Doppler shift is performed using the velocity field computed by TIDES projected along the line-of-sight to the observer.  

{\it Step 4}: Observational data obtained at one orbital phase when the lines of the primary and secondary are well-separated are compared to the synthetic spectra computed for the various sets of \teffp\ and \loggp\ at a similar orbital phase in order to determine the best match between the synthetic and observed photospheric lines. Preference is given to matching lines that are good \teff\ and \logg\ diagnostics\footnote{See, for example, Massey et al. (\cite{Massey09}) for a more in-depth description of the spectrum-matching process.} and, among these, those that lie in spectral regions where the uncertainties in the continuum normalization of the data are minimum.  The atlas of Walborn \& Fitzpatrick (\cite{Walborn}) is used as a guide for the trends in He\,\textsc{i}, He\,\textsc{ii} and Si\,\textsc{iii} line strengths with increasing \teff\ and \logg. For the present investigation, over 160 synthetic model spectra were tested against the observations. A ``best match'' of the synthetic spectrum to the observations is attained when synthetic spectra with adjacent values of \teffp\ and \loggp\ bracket the majority of observed spectral lines amongst the ones used for diagnostics purposes. As will be shown below, our best match synthetic spectra reproduce the observations to within $\sim5$\% of continuum unit. 
 
We initiated the analysis using the orbital elements and stellar parameters derived by Aufdenberg et al. (2008) and the value of ${\mathrm v}\sin i$ from Smith (\cite{Smitha}). The primary reason for this choice is that these parameters were found by Harrington et al. (\cite{Harrington}) to produce  line-profile variability that most resembled the one observed. The qualitative nature of the line profile variations predicted by the TIDES code depends not only on masses, stellar radii and orbital parameters but also on ${\mathrm v}\sin i$, the depth of the layer that is modelled in TIDES, $\Delta R/R_1$, and on the kinematical viscosity, $\nu$ of the material. An adequate combination of all of these parameters is required to achieve a satisfactory match to the observations. For the present analysis we fixed $M_1$, $M_2$, $e$, $\Delta R/R_1$, $\nu$ and the computational parameters required by TIDES to the values that were found by Harrington et al. to best reproduce the line-profile variability. The parameters that were varied are: $R_1$, $R_2$, $i$ (orbital inclination), and the values of \teff\ and \logg\ for the two stars.

After several iterations we found that the best fit to the observations was attained with $R_1^{\rm pole}=6.84 R_{\sun}$ and $i=60\degr$.  Both of these values lie within the uncertainties quoted by Aufdenberg et al. (2008) and hence, we shall refer to the input parameter set as that of Aufdenberg et al. (2008).

We used the ESPaDOns spectra obtained on 22 and 26 March 2008 (orbital phases 0.88 and 0.84, respectively\footnote{The phase $\phi=0$ corresponds to the periastron passage of the observations and we used the T$_0$ and the apsidal period derived by Aufdenberg et al. (2008).}) as the guide for the first synthetic spectrum, since at these orbital phases the absorptions arising in each star  are well separated. The velocity field computed in {\it Step 1} was applied to the grid of models described  in {\it Step 2} and the synthetic spectra ({\it Step 3}) were compared with these ESPaDOns spectra.

\begin{figure}
\resizebox{\hsize}{!}{\includegraphics{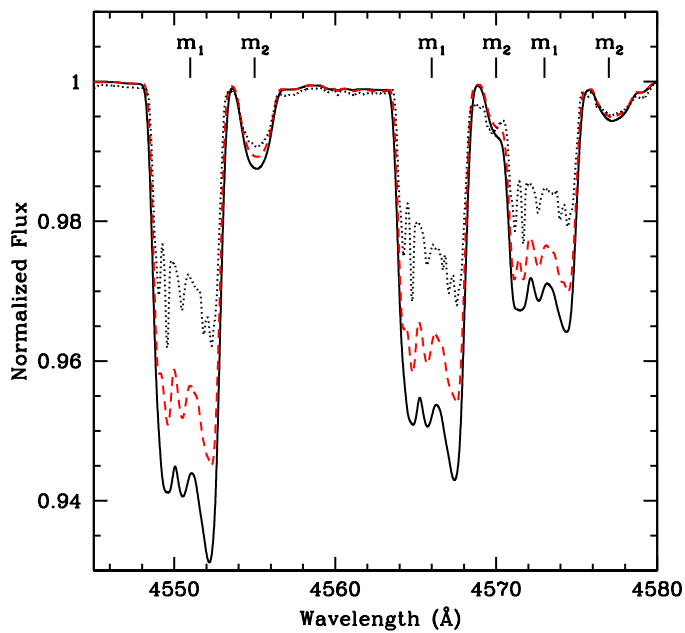}}
\caption{Synthetic Si\,\textsc{iii} line profiles computed for {\it Spica} with the TLUSTY stellar atmosphere models for \teff\ and \logg\ appropriate to each surface element and Doppler-shifted using the surface velocity field computed with TIDES.
The three spectra correspond to microturbulent velocities ${\mathrm v}_{turb}= 2$ (dots), 10 (dashes) and 15 (continuous) \kms\ and illustrate how the larger values of this  parameter produce greater line-strength in the \si{iii}{4552} line. The lines originating in the primary and secondary are indicated with  $m_1$, and $m_2$ respectively. The depicted profiles correspond to orbital phase $\phi=0.75$.
\label{fig_evolution_vturb}}
\end{figure}

\begin{figure}
\resizebox{\hsize}{!}{\includegraphics{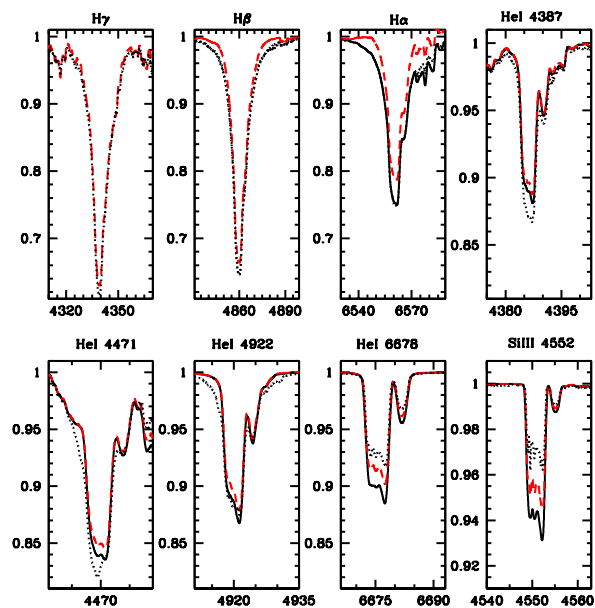}}
\caption{Dependence of line strength on microturbulent speed ${\mathrm v}_{turb}$. As in Fig.~\ref{fig_evolution_vturb}, the same stellar atmosphere model computed with ${\mathrm v}_{turb}= 2$ (dots), 10 (dashes) and 15 (continuous) \kms\ is shown. The behaviour of the stronger He\,\textsc{i} lines is different from that of weaker lines. In the H-lines, the dependence on ${\mathrm v}_{turb}$ is strongest in the line wings.
\label{panels_vturb}}
\end{figure}

In the initial iteration for finding the optimum values of \teffp, \loggp\ an excellent match was attained for the H and He\,\textsc{i} lines with ${\mathrm v}_{turb}=2$ \kms, but  the majority of the heavy element lines were too weak by factors of 2--3. Thus, the next iteration consisted of finding the value of ${\mathrm v}_{turb}$ for which the strength of the Si\,\textsc{iii} $\lambda\lambda$\,4552-4574 triplet lines coincided with the observations\footnote{An in-depth description of the manner in which ${\mathrm v}_{turb}$ is generally determined may be found in Hunter et al. (\cite{Hunter}). Included is a justification of the use of the Si\,\textsc{iii} triplet for fixing the value of this parameter as well as a discussion of the manner in which the use of different lines results in different values.}. We found that a good match to the Si\,\textsc{iii} line strengths requires $10\leq {\mathrm v}_{turb}/(\text{\kms})\leq15$, and the final model was computed with ${\mathrm v}_{turb}=15$ \kms. It must be noted, however, that this value of ${\mathrm v}_{turb}$ resulted in lines such as O\,\textsc{ii}\,$\lambda\lambda$\,4072-79 being significantly stronger in the synthetic spectra than in the observations. Fig.~\ref{fig_evolution_vturb} illustrates the manner in which the Si\,\textsc{iii} triplet changes for ${\mathrm v}_{turb}=2$, 10 and 15 \kms, and Fig.~\ref{panels_vturb} shows the effect on other lines. Noteworthy is the very different behaviour of \he{i}{4471} from that of \si{iii}{4552} and other heavy-element lines. This difference may be traced to the atomic line-strength parameters which affect the shape of the line profiles.

Once the set of parameters that best described the ESPaDOnS spectrum of 22 March were determined, these were held constant and we proceeded to compute the synthetic spectra at 20 equally spaced orbital phases. These are the spectra that are discussed in the next sections.

\section{Results}

\subsection{The impact of interaction effects on \teff\ and \logg}

 The best match to the ESPaDOns spectra of 22 and 26 March was attained with (\teffp, \loggp) of ($24000\pm500$ K, 3.78) for the primary and ($19500\pm500$ K, 4.16) for the secondary. These values are in good agreement with those found by Aufdenberg et al. (2008) for the primary star, ($24750\pm500$, $3.71\pm0.06$). For the secondary, our result is in agreement with Aufdenberg et al.'s \loggp$=4.16\pm0.05$, and Lyubimkov et al. (\cite{Lyubimkov}) \teffp$=20800\pm1500$ K.   

\begin{figure}
\resizebox{\hsize}{!}{\includegraphics{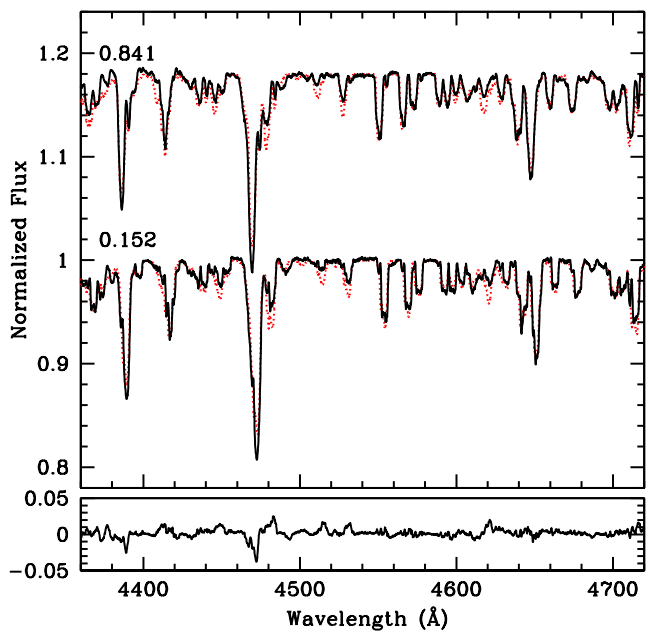}}
\caption{Comparison of the best-fit CoMBiSpeC model (dots) with the ESPaDOns spectra of {\it Spica} at orbital phases 0.152 (15 March)  and 0.841 (26 March; shifted by +0.2 continuum units). The tracing shown at the bottom is the difference between the 15 March spectrum and its corresponding synthetic spectrum and shows that the fit is good to $\sim3$\%.
\label{fig_spectra_general1}}
\end{figure}

\begin{figure}
\resizebox{\hsize}{!}{\includegraphics{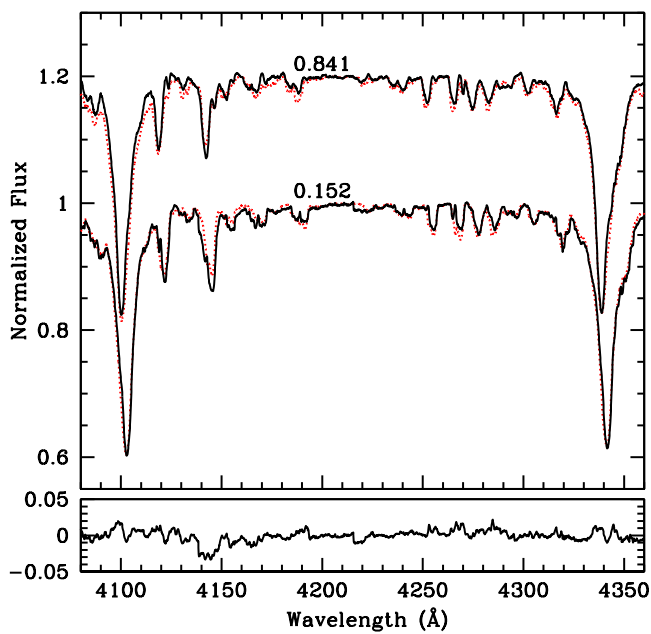}}
\caption{Same as previous figure for the spectral region containing H$\delta$ and H$\gamma$.
\label{fig_spectra_general3}}
\end{figure}

\begin{figure}
\resizebox{\hsize}{!}{\includegraphics{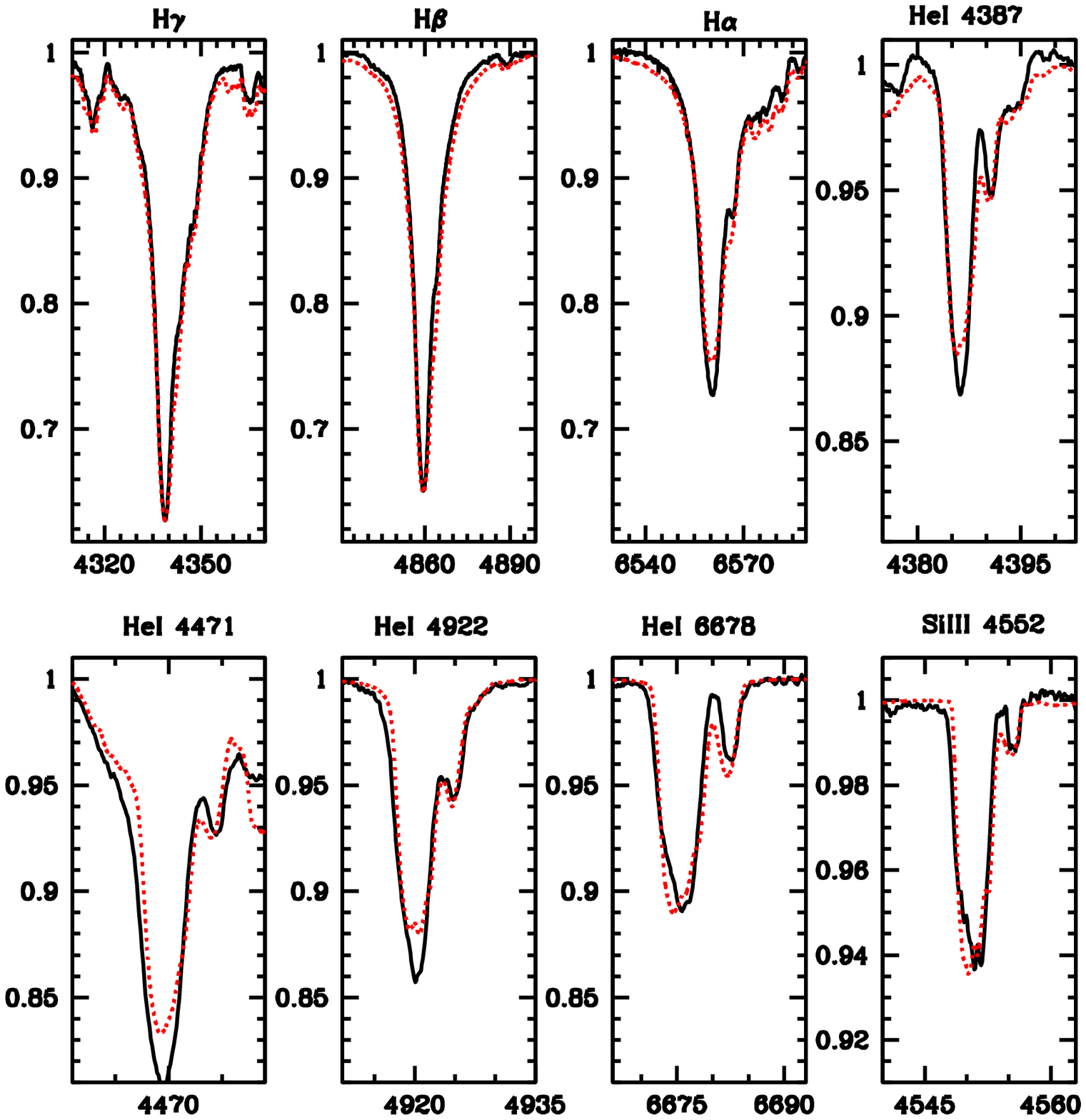}}
\caption{Individual lines in the 26 March spectrum compared to the corresponding synthetic spectrum.
\label{fig_panels_26march}}
\end{figure}

\begin{figure}
\resizebox{\hsize}{!}{\includegraphics{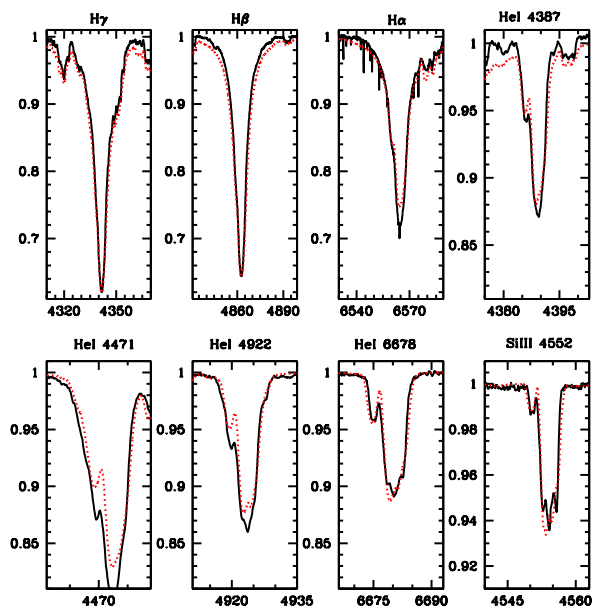}}
\caption{Individual lines in the 28 March spectrum compared to the corresponding synthetic spectrum.
\label{fig_panels_28march}}
\end{figure}

The parameters of the final model are listed in Table~\ref{modelprop}, and selected spectral regions of these models and the corresponding observational data of 26 March ($\phi=0.84$) and 15 March ($\phi=0.15$) are shown in Figs.~\ref{fig_spectra_general1} and \ref{fig_spectra_general3}. Figs.~\ref{fig_panels_26march} and \ref{fig_panels_28march} illustrate the matching between observed and synthetic spectra for individual lines. The difference between the observed and the synthetic spectra is $\leq5$\%, as shown by the plot at the bottom of these figures. For the 4060-4360 \AA\ and 4360-4720 \AA\ wavelength regions, the maximum difference is $<$4\% except at $\lambda$\,4078.4 (8.7\%). For the 4730-5130 \AA\ wavelength region, the maximum difference is 2.5\% and appears in the wings of H$\beta$.  

\begin{table}[ht]
	\begin{center}
	 \caption{Parameters used for computation with the TIDES + CoMBiSpeC model.\label{modelprop}}
		\begin{tabular}{l c c}
			\hline\hline
			Parameters & Primary & Secondary \\
			\hline\hline
			Common parameters &  & \\
			\hline
			Period (day) & \multicolumn{2}{c}{$4.01452$} \\
			Eccentricity & \multicolumn{2}{c}{$0.067$} \\
			$\omega^{\text{(a)}}$ ($\degr$) & \multicolumn{2}{c}{$255$} \\
			Inclination ($\degr$) & \multicolumn{2}{c}{$60$} \\
			\hline\hline
			CoMBiSpeC parameters &  & \\
			\hline	
			Mass ($M_{\sun}$)& $10.25$ & $6.97$ \\
			Polar temperature (K) & $24000$ & $19500$ \\
			Polar radius ($R_{\sun}$) & $6.84$ & $3.64$ \\
			Polar log(g) (cgs) & $\simeq 3.78$ & $\simeq 4.16$ \\
			${\mathrm v}_{rot}$ (\kms) & $199$ & $87$ \\
			${\mathrm v}\sin i$ (\kms) & $172$ & $75$\\
			$\beta_0$ & $2.07$ & $1.67$\\
			Microturbulent velocity (\kms) & $15$ & $15$\\
			\hline\hline
			TIDES code parameters &  & \\
			\hline
			Viscosity, $\nu$ ($R_{\sun}^2$day$^{-1}$)$^{\text{(b)}}$ & $0.028$ & $0.028$ \\
			Layer depth ($\Delta R/R$) & $0.07$ & $0.07$\\
			Polytropic index & $1.5$ & $1.5$ \\
			Number of azimuthal partitions & $500$ & $500$\\		
			Number of latitudinal partitions& $20$ & $20$\\	
	  	\hline
		\end{tabular}
		\end{center}
 	\tablefoot{$^\text{(a)}$ Argument of periastron of the secondary. $^\text{(b)}$ 1 R$_{\sun}^2$day$^{-1}$= 5.67$\times$10$^{16}$ cm$^2$s$^{-1}$.}
	\end{table}

\begin{figure}
\resizebox{\hsize}{!}{\includegraphics{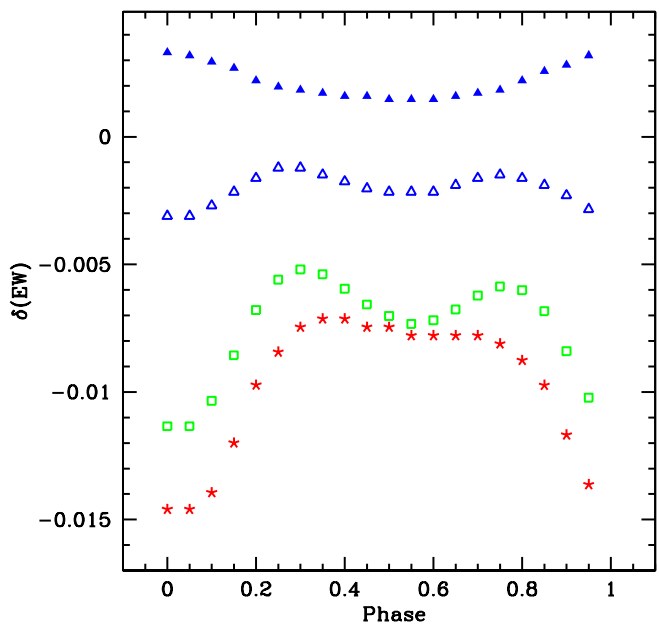}}
\caption{The difference between equivalent widths measured on the perturbed and unperturbed spectra, $\Delta$(EW)=[EW$_{p}$-EW$_{u}$]/EW$_{u}$, of the primary star in {\it Spica} showing that the effects are $\leq1.5$\%, and thus indicating that the interaction effects have a negligible impact on the spectral classification. Negative residuals indicate that the absorption in the perturbed profiles is weaker than in the unperturbed profiles. The different symbols correspond to: \he{i}{4921}(open triangles), \he{i}{5875} (filled-in triangles), \si{iii}{4552} (stars) and H$\beta$ (squares).
\label{fig_EWs}}
\end{figure}

The maximum departure from sphericity in {\it Spica}'s primary is $\leq0.2 R_{\sun}$, and thus the value of \logg\ over the stellar surface is practically constant. The largest \teff\ is found at the pole and decreases towards the equator, but the small gravity darkening leads to a \teff\ decrease of only $\sim500$\,K. The irradiation effect is also small and thus the hemisphere facing the secondary is nearly at the same temperature as the opposite hemisphere. The small differences in \teff\ and \logg\ over the stellar surface in the primary lead to line strengths in the perturbed spectra that differ only slightly from those in the unperturbed spectrum. Fig.~\ref{fig_EWs} shows that the differences in the equivalent width (EW) of the lines in the perturbed and unperturbed spectra are $\leq1.5$\%, which is below the typical observational uncertainties and those inherent to the spectrum fitting technique. Thus we conclude that despite the strong line-profile variability in {\it Spica}'s primary star, there is a negligible effect on the derivation of \teff\ and \logg\ values.

\begin{figure}
\resizebox{\hsize}{!}{\includegraphics{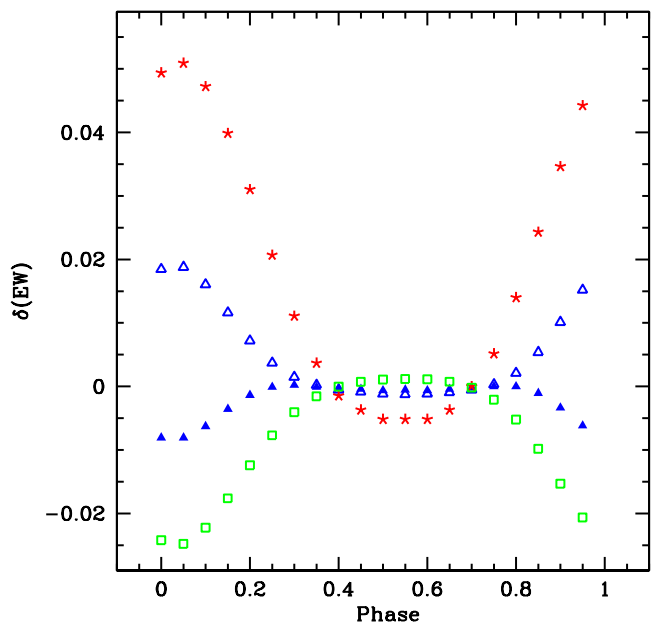}}
\caption{Same as previous figure (Fig.~\ref{fig_EWs}) for the secondary showing that in this case, the  effects are $\leq5$\%. Note that around periastron, the \si{iii}{4552} strength increases by $\sim5$\% while that of H$\beta$ decreases by $\sim2.5$\%.
\label{fig_EWs_sec}}
\end{figure}

For the secondary, the effects caused by gravity darkening are also small, but due to the hotter primary the irradiation effects are more important.  Hence, the largest \teff$\sim 20300$\,K is on the equator of the hemisphere facing the primary while the lowest \teff$=19400$\,K occurs on the opposite hemisphere, also along the equator. The irradiated hemisphere of the secondary is viewed by the observer around periastron passage, at which time the EW of \si{iii}{4552} increases by $\sim5$\% while that of H$\beta$ decreases by $\sim2$\%, as illustrated in Fig.~\ref{fig_EWs_sec}. 

It is interesting to note that despite the relatively small surface deformations, the observed light curve of {\it Spica} displays ellipsoidal variations (Shobbrook et al. \cite{Shobbrook1}, Sterken et al. \cite{Sterken}) with peak-to-peak amplitudes $\sim 0.03$ mag. We computed the predicted light curve variations from out model results following the method described in Palate \& Rauw (\cite{Palate}) and find a peak-to-peak amplitude $\sim0.02$ mag. The difference between our predicted light curve and that which is observed may in part be due to the uncertainties introduced by the short-period photometric variability which is present in the observations.

\subsection{Line profile variability}

\begin{figure}
\resizebox{\hsize}{!}{\includegraphics{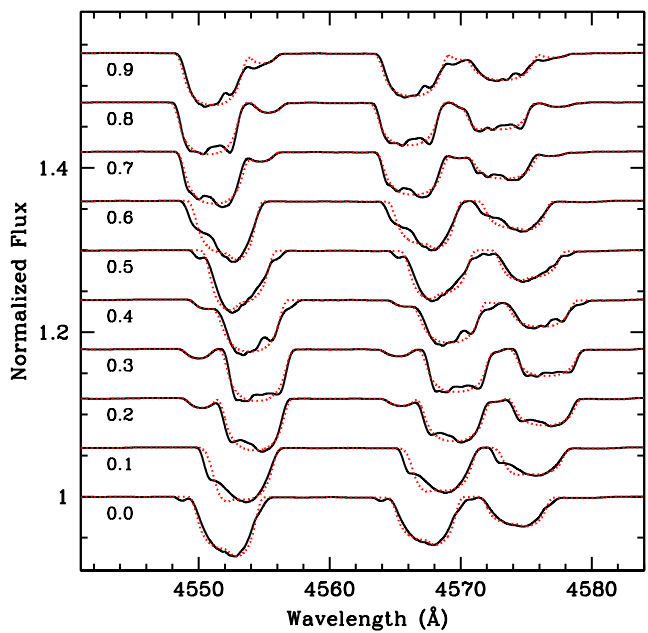}}
\caption{The synthetic  primary + secondary combined spectra of the Si\,\textsc{iii} triplet are stacked in order of increasing orbital phase ($\phi$=0 corresponds to periastron). The perturbed spectra are displayed with the dark tracing and the unperturbed spectra with dots. The ``bumps'' in the primary star's perturbed profiles are evident as is the difficulty they introduce in properly locating the contribution of the secondary except at $\phi \sim0.3\pm0.05$ and $0.8\pm0.05$ when the contribution from the secondary is clearly resolved.
\label{fig_montage_model}}
\end{figure}

\begin{figure}
\resizebox{\hsize}{!}{\includegraphics{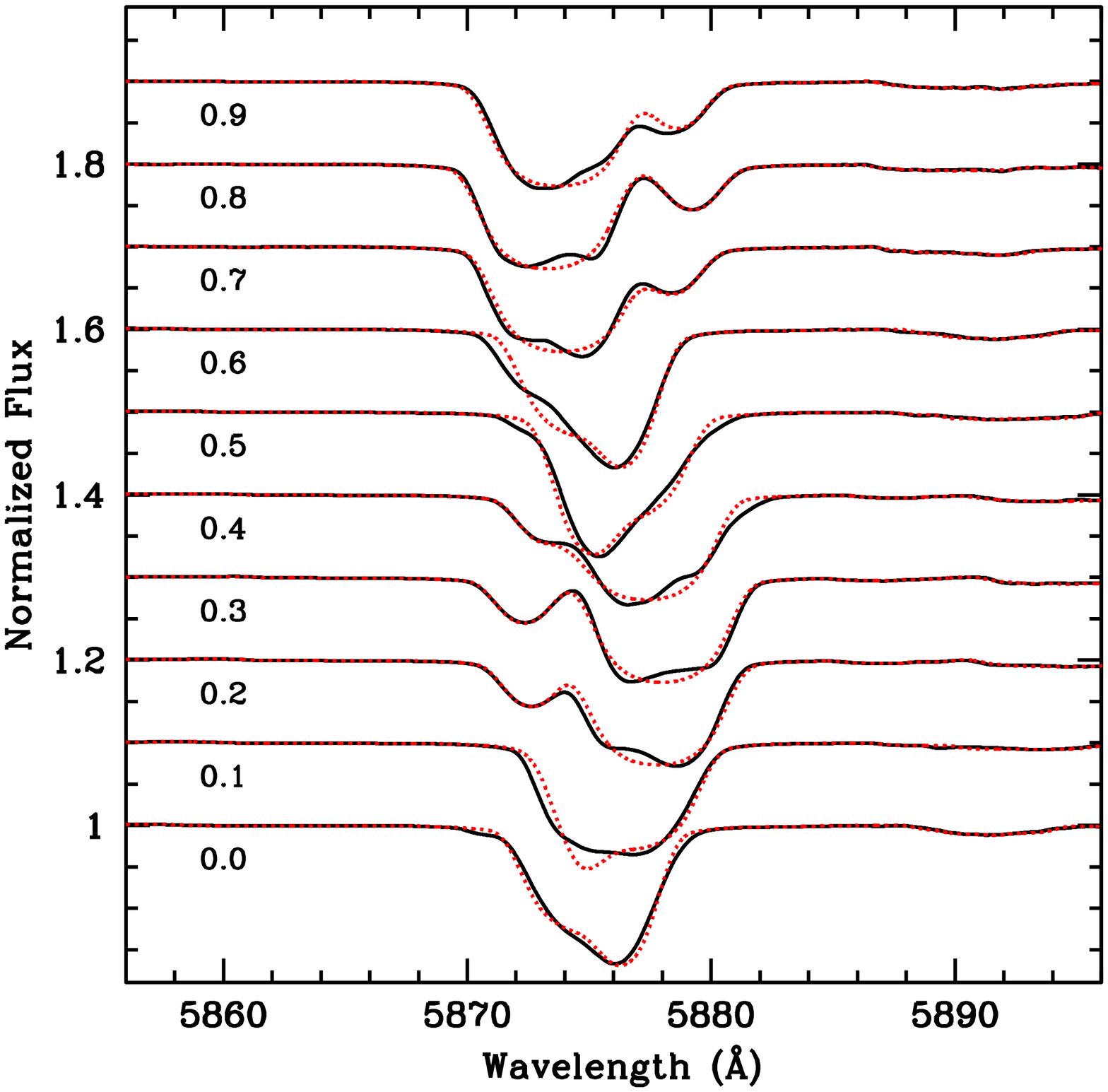}}
\caption{Same as previous figure for \he{i}{5875}. 
\label{fig_montage_model_5875}}
\end{figure}

The binary interactions  significantly affect the shape of the line profiles. Fig.~\ref{fig_montage_model} shows the Si\,\textsc{iii} triplet line profiles at 10 orbital phases in the perturbed and the unperturbed spectra. The strong phase-dependent variations in the perturbed profiles are clearly seen and can be described primarily in terms of ``bumps'' and asymmetries, similar to those present in the observational data\footnote{See Harrington et al. \cite{Harrington} where a set of spectra computed at very short time steps and showing the moving ``bumps'' is illustrated.}. The same behaviour is present in numerous other photospheric absorptions, such as \he{i}{5875} (Fig.~\ref{fig_montage_model_5875}), but is at the $\sim1.5$\% level. 

\begin{figure}
\resizebox{\hsize}{!}{\includegraphics{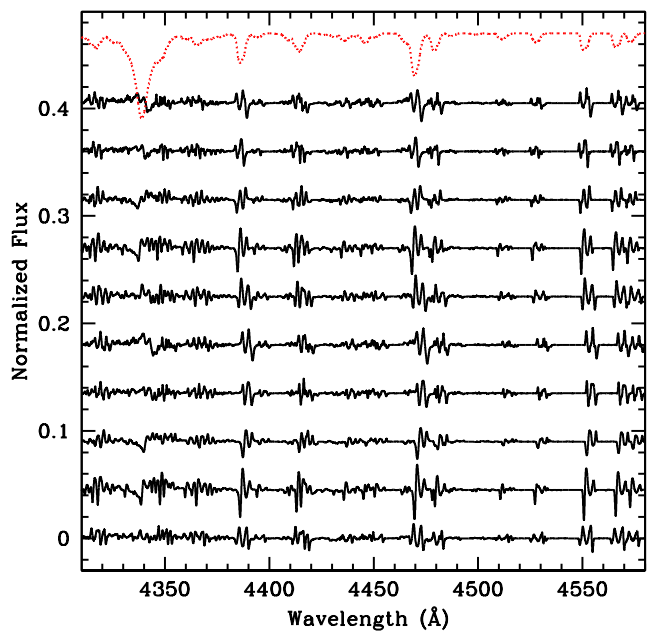}}
\caption{Difference between the synthetic perturbed spectra of the primary and the corresponding unperturbed spectra showing that the lines of intermediate intensity display the most prominent variations in the ``bump'' structure. The orbital phases increase from bottom to top 
with increments of $\Delta\phi=0.1$ and starting with $\phi=0$. Differences are shifted along the vertical axis for clarity. The dotted tracing at the top is the perturbed spectrum at $\phi=0.90$, scaled and shifted to fit in this figure.  
\label{fig_ratio_model_1}}
\end{figure}

The most prominent  perturbations occur in absorption lines of intermediate intensity, as is shown in Fig.~\ref{fig_ratio_model_1}, where we plot the difference between the perturbed and  unperturbed synthetic spectra at 10 phases in the orbital cycle for the $\lambda\lambda$\,4310-4580 \AA\ region. Strong lines, such as H$\gamma$, undergo weaker perturbations than lines such as \he{i}{4471} and the Si\,\textsc{iii} triplet.

\subsection{Distortion of the RV curve}

The line-profile variations introduce an intrinsic uncertainty in the radial velocity (RV) measurements which, as will be shown below, leads to systematic distortions in the RV curve. To illustrate this point, the centroid of the lines \he{i}{4921} and $\lambda$\,5875, \si{iii}{4552} and H$\beta$ were measured in the  perturbed and unperturbed spectra by numerical integration between two fixed positions at the continuum level on both sides of the absorption minimum.\footnote{The functional form is $\frac{\Sigma x_i (1-y_i)}{\Sigma (1-y_i)}$, where $x_i$, $y_i$ are the values of wavelength and flux at each wavelength step. The wavelength step of our synthetic spectra is 0.01 \AA\ and integration intervals typically contained $\sim2000$ points.}

\begin{figure}
\resizebox{\hsize}{!}{\includegraphics{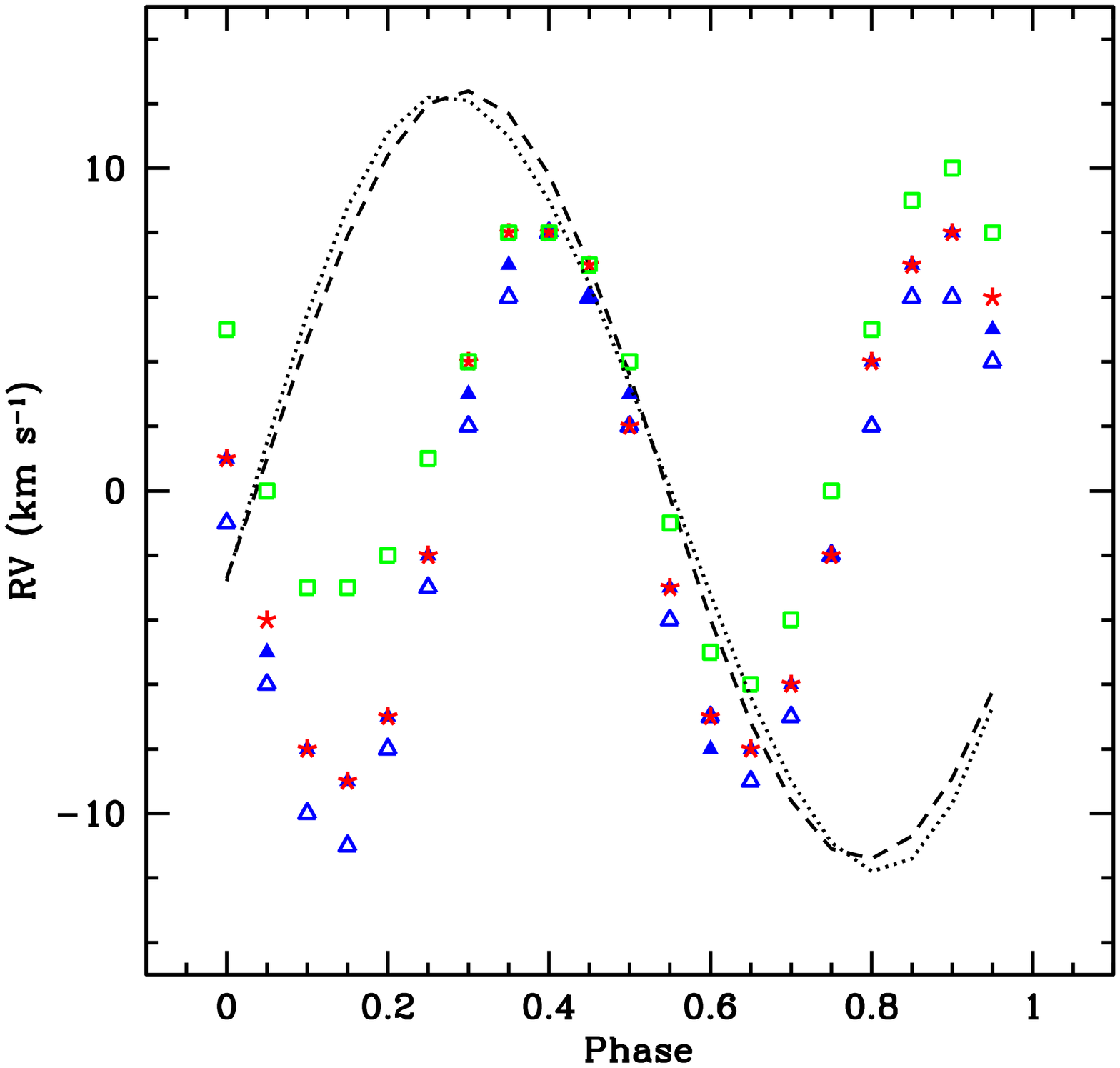}}
\caption{The difference between RV measurements made on the perturbed and unperturbed spectra $\delta\text{(RV)}=\text{(RV)}_{p}-\text{(RV)}_{u}$ for \he{i}{4921} (open triangles), \he{i}{5875} (filled-in triangles), \si{iii}{4552} (stars)  and H$\beta$ (squares).  Also shown is the shape of the perturbed (dash) and unperturbed (dots)  RV curves for \he{i}{5875} after down-scaling by a factor of 10 for illustration purposes. The perturbation leads to a RV curve corresponding to a more eccentric orbit than the actual orbit. 
\label{fig_RVs}}
\end{figure}

Fig.~\ref{fig_RVs} shows the difference $\delta\text{(RV)}=\text{(RV)}_{p}-\text{(RV)}_{u}$ for the primary, where the subscripts 'p' and 'u' represent the perturbed and unperturbed spectra, respectively, and the RV$_u$ correspond to the actual orbital motion. The maximum semi-amplitudes of $\delta\text{(RV)}$ range between 6 and 10 \kms, depending on the particular line being measured. A similar analysis performed using a cross-correlation method\footnote{The FXCOR routine in the {\it Image Reduction Analysis Facility (IRAF)} package.} yields a semi-amplitude $\delta\text{(RV)} \sim7$ \kms\ for the two wavelength intervals that were cross-correlated ($\lambda\lambda~4150-4300$ \AA\ and $\lambda\lambda~5200-5800$ \AA), consistent with the result obtained for individual lines. The shape of the perturbed and unperturbed RV curves for \he{i}{5875} is also shown in Fig.~\ref{fig_RVs} (after scaling by a factor of 10 for illustration purposes) showing that the strongest distortion occurs on the ascending and descending branches of the RV curve, on both sides of the extrema; i.e., at  orbital phases $\sim0.15$, 0.40, 0.65 and 0.90. The combined effect of these distortions is to skew the RV curve, giving it the appearance of one with a larger eccentricity than that of the actual orbit.

\begin{figure}
\resizebox{\hsize}{!}{\includegraphics{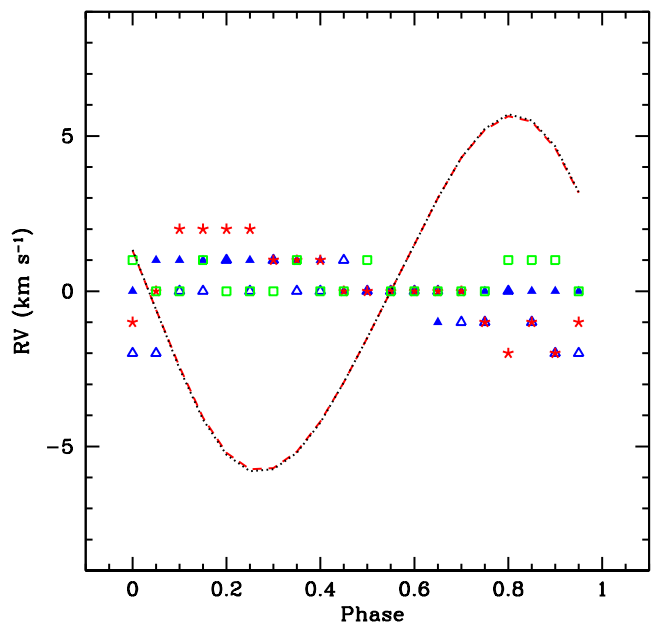}}
\caption{As in the previous figure, the difference between RV measurements made on the perturbed and unperturbed spectra are plotted but here for the secondary star, and the scaled RV curves are those of \si{iii}{4552}, which shows the strongest perturbation effects.
\label{fig_RVs_sec}}
\end{figure}

For the secondary star, its weaker line-profile variability leads to much smaller deformations in the RV curves (see Fig.~\ref{fig_RVs_sec}), $\delta$(RV)$=2$ \kms, with extrema at $\phi \sim0.2$ and 0.85. The most strongly perturbed RV curve is that of \si{iii}{4552} due to the greater sensitivity of this line to the heating by irradiation of the companion.

\subsection{Comparison of CoMBiSpeC and TIDES model line profiles}

\begin{figure}
\resizebox{\hsize}{!}{\includegraphics{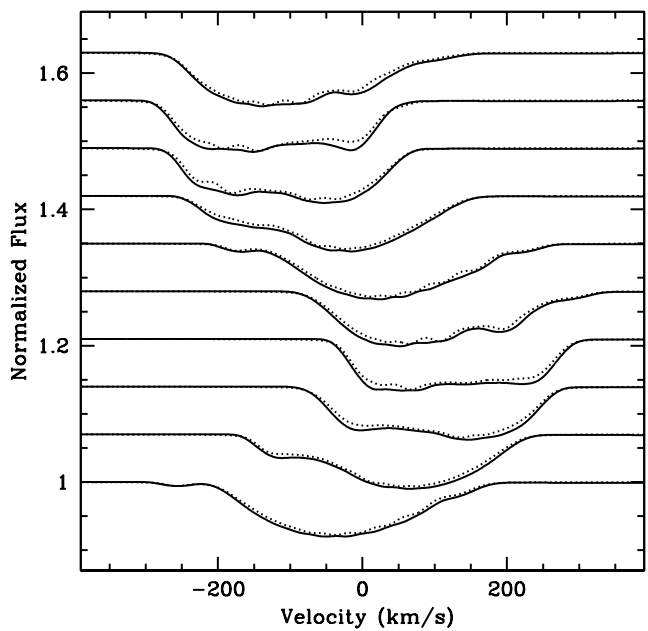}}
\caption{The \si{iii}{4552} line profiles computed with ComBiSpeC compared with the line profiles computed with the TIDES code (dots). In this TIDES calculation, the computation is performed using Voigt local line profiles. The ComBiSpeC code uses the same velocity field that is computed with TIDES, but the local line profile is obtained from the non-LTE TLUSTY radiative transfer calculation. Only the profiles computed for the primary star are shown here.
\label{fig_compare_TIDES}}
\end{figure}

The CoMBiSpeC calculation uses the velocity field computed by TIDES to compute the line profiles. A line-profile calculation is also implemented in TIDES (Moreno et al. \cite{Moreno}) which in the original version employed in Harrington et al. (\cite{Harrington}) 
assumed a Gaussian local line profile instead of a line profile generated from a stellar atmosphere calculation.  In Fig.~\ref{fig_compare_TIDES} we compare the line profiles produced by both calculations, but in this case TIDES uses a Voigt shape for the local line profiles with a Lorentzian coefficient $a_L=0.3$. Both calculations use the same surface velocity field (calculated by TIDES) and microturbulent 
speeds of 15 \kms, and all the binary parameters are the same. The differences between the TIDES and the CoMBiSpeC calculations are minimal.

Because TIDES computes the line profile variability very rapidly, its use is desirable for the analysis of a large parameter space such as that involving stellar radius, rotation velocity, eccentricity, orbital inclination, layer depth and viscosity,  and the above justifies its use for these purposes.  

\section{Summary and Conclusions}

In this paper we explore the uncertainties that binary interaction effects may introduce in the derivation of fundamental stellar parameters when using stellar atmosphere models constructed for single stars. Our test case is {\it Spica}, a relatively close B-type binary system with unevolved, detached components. We use the TIDES code to compute the surface deformation and velocity field, the TLUSTY model atmosphere synthetic spectral grids, and the CoMBiSpeC model to synthesize output spectra that include the effects of irradiation by the companion and the tidal perturbations. These synthetic spectra are compared to high {\it S/N} and high resolution observations from which the following results are derived:

\begin{enumerate}
\item{} The phase-dependent variations in line strength (EW) due to the interactions is $\leq2$\% for the primary and $\leq4$\% for the secondary, leading to values of \teff\ and \logg\ that are indistinguishable, within the uncertainties, from those derived from models that neglect the interactions. The reason for this is that the irradiation effects and the departure from sphericity of both stellar surfaces are very small.  The radius of the secondary is $\sim7$ times smaller than the orbital separation, so even though the primary is hotter, this large separation guarantees that only a small irradiation effect is present. Similar considerations apply to the tidal deformations.  The radius of the primary is only $\sim3.6$ times smaller than the orbital separation, but the irradiation from the secondary is insignificant due to its cooler temperature and, due to it's smaller mass, the tidal deformations are also negligible. 

\item{} The primary star rotates super-synchronously, which significantly perturbs its surface, leading to strong phase-dependent line profile variations. As a result, the radial velocity curve is distorted with respect to the curve that describes the orbital motion, with maximum deviations of $\leq10$ \kms.  Although the peak-to-peak amplitude of the RV curve is not affected, the shape is skewed so that a larger eccentricity than the actual value is inferred. We note that this effect may be the source of the discrepancy between the values of $e$ that are given by Riddle ($e=0.067$) and Herbison-Evans et al. ($e=0.146$).

\item{} The velocity structure that is computed by TIDES leads naturally to the presence of bumps on the profiles of lines such as the Si\,\textsc{iii}\,$\lambda\lambda$\,4552-72 triplet and other lines of intermediate strength without the need of any {\it ad hoc} assumption regarding non-radial pulsations. The nature of the bumps in the synthetic spectra is qualitatively similar to that in the observations for all lines contained in our spectra.

\item{} The weaker lines in the spectrum require ${\mathrm v}_{turb}=10-15$ \kms, whereas He\,\textsc{i} lines such as $\lambda$\,4471 and O\,\textsc{ii} favour a smaller value. Interestingly, the lines that show the clearest bumps are the ones that require the larger values of ${\mathrm v}_{turb}$ to attain an adequate match to the line-strengths.  
 
\end{enumerate}

We conclude that for a binary system such as {\it Spica}, the uncertainties in the model-fitting process (NLTE effects, microturbulence, rectification of the continuum level in the observations) are larger than those introduced by the tidal velocity field and heating effects. Future work will extend the exploration introduced in this paper to systems with parameters that are different from {\it Spica}'s, where irradiation and tidal deformations are more important, in order to evaluate the uncertainties that are introduced in these cases
with the use of single-star atmosphere models.

The above being said, it must be noted that nearly all of the synthetic model atmospheres are  computed under the assumptions of hydrostatic and radiative equilibrium.  However, the shear produced by different surface layers as they slide with respect to each other in response to the tidal perturbation leads to energy dissipation (Toledano et al. \cite{Toledano}; Moreno et al. \cite{Moreno}). Depending on the particular regions where this energy is deposited, the temperature structure of the outer layers may be altered with respect to that derived when radiative equilibrium is strictly enforced. Furthermore, the dynamical nature of the stellar photosphere could influence the assumptions that are used for fixing ${\mathrm v}_{turb}$, for example, assuming that it is constant over the stellar surface. A hint pointing towards a non-constant value of ${\mathrm v}_{turb}$ is suggested by our results. Thus, the applicability of single-star models to close binary stars may need to be critically assessed.

\begin{acknowledgements}
We wish to express our gratitude to Ivan Hubeny for guidance in the use and implementation of SYNSPEC49 and Andres Sixtos for performing the IRAF cross-correlation computation of the RVs. The authors also would like to acknowledge Jason Aufdenberg for a careful reading of this paper and helpful comments. We acknowledge support through the XMM/INTEGRAL PRODEX contract (Belspo), from the Fonds de Recherche Scientifique (FRS/FNRS); UNAM/DGAPA/PAPIIT grant IN-105313, and CONACYT grant 129343.
\end{acknowledgements}

\end{document}